\pdfoutput=1
\documentclass[12pt]{iopart}
\usepackage{iopams}  
\usepackage{bm}
\usepackage{graphicx}
\begin{document}

% ******************************************************************
\newcommand{\unit}[1]{\ensuremath{~\mathrm{#1}}}
\newcommand{\unitnospace}[1]{\ensuremath{\mathrm{#1}}}
\newcommand{\K}{\unit{K}}
\newcommand{\T}{\unit{T}}
\newcommand{\mT}{\unit{mT}}
\newcommand{\uT}{\unit{\mu{}T}}
\newcommand{\A}{\unit{A}}
\newcommand{\nA}{\unit{nA}}
\newcommand{\pA}{\unit{pA}}
\newcommand{\kV}{\unit{kV}}
\newcommand{\uV}{\unit{\mu{}V}}
\newcommand{\mm}{\unit{mm}}
\newcommand{\um}{\unit{\mu{}m}}
\newcommand{\nm}{\unit{nm}}
\newcommand{\degree}{\unitnospace{^\circ}}

\newcommand{\E}[1]{\ensuremath{\!\times\! 10 ^{#1}}}

	% with correct space after "."
\newcommand{\vs}{vs\xspace}						% without "."
\newcommand{\ie}{i.e.\ }							% with correct space after "."
% ******************************************************************

% Journal identifier can be put here if required, e.g.
\SUST

\title[Flux-transfer losses in helically wound superconducting power cables]{Flux-transfer losses in helically wound superconducting power cables}

\author{John R Clem$^1$and A P Malozemoff$^2$\footnote[3]{To whom correspondence should be
addressed.} }

\address{$^1$\ Ames Laboratory and Department of Physics and Astronomy,\\
  Iowa State University, Ames, Iowa, 50011--3160, USA}

\address{$^2$\ AMSC, 64 Jackson Road, Devens, MA 01434, USA}

 \ead{amalozemoff@amsc.com}

\begin{abstract}Minimization of ac losses is essential for economic operation of high-temperature superconductor (HTS) ac power cables.  A favorable configuration for the phase conductor of such cables has two counter-wound layers of HTS tape-shaped wires lying next to each other and helically wound around a flexible cylindrical former.  However, if magnetic materials such as magnetic substrates of the tapes lie between the two layers, or if the winding pitch angles are not opposite and essentially equal in magnitude to each other, current distributes unequally between the two layers.  Then, if at some point in the ac cycle the current of either of the two layers exceeds its critical current, a large ac loss arises from the transfer of flux between the two layers.  A detailed review of the formalism, its application to the case of paramagnetic substrates including the calculation of this flux transfer loss  is presented.

\end{abstract}


\submitted{\SUST}

% Comment out if separate title page not required
\maketitle

\section{Introduction}

	Reducing ac losses is critical to successful design of ac power transmission cables based on helically wound high-temperature superconductor (HTS) tapes \cite{Kalsi11}.  The tapes, typically of order 100 $\mu$m thick, are wound around a long flexible cylindrical former typically 2-3 cm in diameter.  The theory of ac losses in such a configuration has been developed by many authors \cite{Garber76,Campbell99,Mawatari08,Amemiya07,Malozemoff09,Clem10}, and numerous loss mechanisms have been identified.  We consider here a particularly large loss mechanism, called flux-transfer loss, which arises in multilayer helically wound cables when current flow is imbalanced between the different layers.
	
Such imbalance is extreme in the case of nested cylinders since as current increases during an ac cycle, it always starts flowing on the outer layers initially and then transfers to the next layer once the critical current of the outer layer has been reached.  The flux transfer between layers gives rise to the flux-transfer loss, a hysteretic loss mechanism which can be understood on the basis of the Bean critical state theory \cite{Bean62,Campbell72,Daeumling99,Daeumling04}.  This loss is much larger than the hysteretic loss within the superconductor cylinders if the cylinder thicknesses are small compared to the space between them.
	
In helically wound designs, a judicious choice of winding pitch generates mutual inductances which can balance the currents in the different layers \cite{Kalsi11,Clem10,Mukoyama97}.  Nevertheless this balancing becomes increasingly difficult to do as the number of layers increases beyond two, and if magnetic or even paramagnetic material lies between the two layers, for example in the substrate of the superconductor tape, balancing becomes impossible.  Flux-transfer losses become important once the current in one of the layers reaches its critical current.

Here this flux-transfer loss is derived for the case of two helically wound layers of second generation coated conductor tapes, as in the configuration widely used for power cables \cite{Kalsi11}.  In such tapes the superconductor film is typically only 1 $\mu$m thick, while the spacing between the two winding layers is of order 100 $\mu$m or more.   In this case one can neglect the hysteretic losses within the superconductor layers  compared to the much larger loss arising from the flux transfer between the layers.  Furthermore, as long as the edges of neighboring tapes in each winding layer are close to each other, losses arising from the finite gaps between the tapes are also small in comparison, and we ignore them as well as the  ``polygonal" losses arising from corners between abutting flat tapes.  A key result of the calculation has been reported earlier \cite{Clem10}, but the full derivation is given here, along with a detailed review of the underlying equations and further discussion of the significance of the flux-transfer loss.

If the substrates of the superconductor tapes are magnetic, the results depend strongly on the substrate orientations relative to the superconductor films.  For future reference, we define {\it in-out} to refer to the two-layer configuration in which the first or inner layer has its substrate facing inward toward the center of the cable, relative to its superconductor film, while the second or outer layer has its substrate facing outward.  Similarly the {\it out-out} configuration has the substrates of both layers facing out relative to the superconductor of each layer, and so on.  The {\it out-out}, {\it in-in}, and {\it out-in} configurations all have magnetic material between the two superconductor layers.  Only the {\it in-out} configuration avoids magnetic material between the superconductors, and as has been pointed out earlier \cite{Clem10}, this is the most favorable configuration to avoid large flux-transfer losses as the cable current approaches its overall critical current.

\section{$N$ layers\label{Nlayers}}

\subsection{Voltages}

\begin{figure}%***** Fig.1 ************************
\includegraphics[width=6cm]{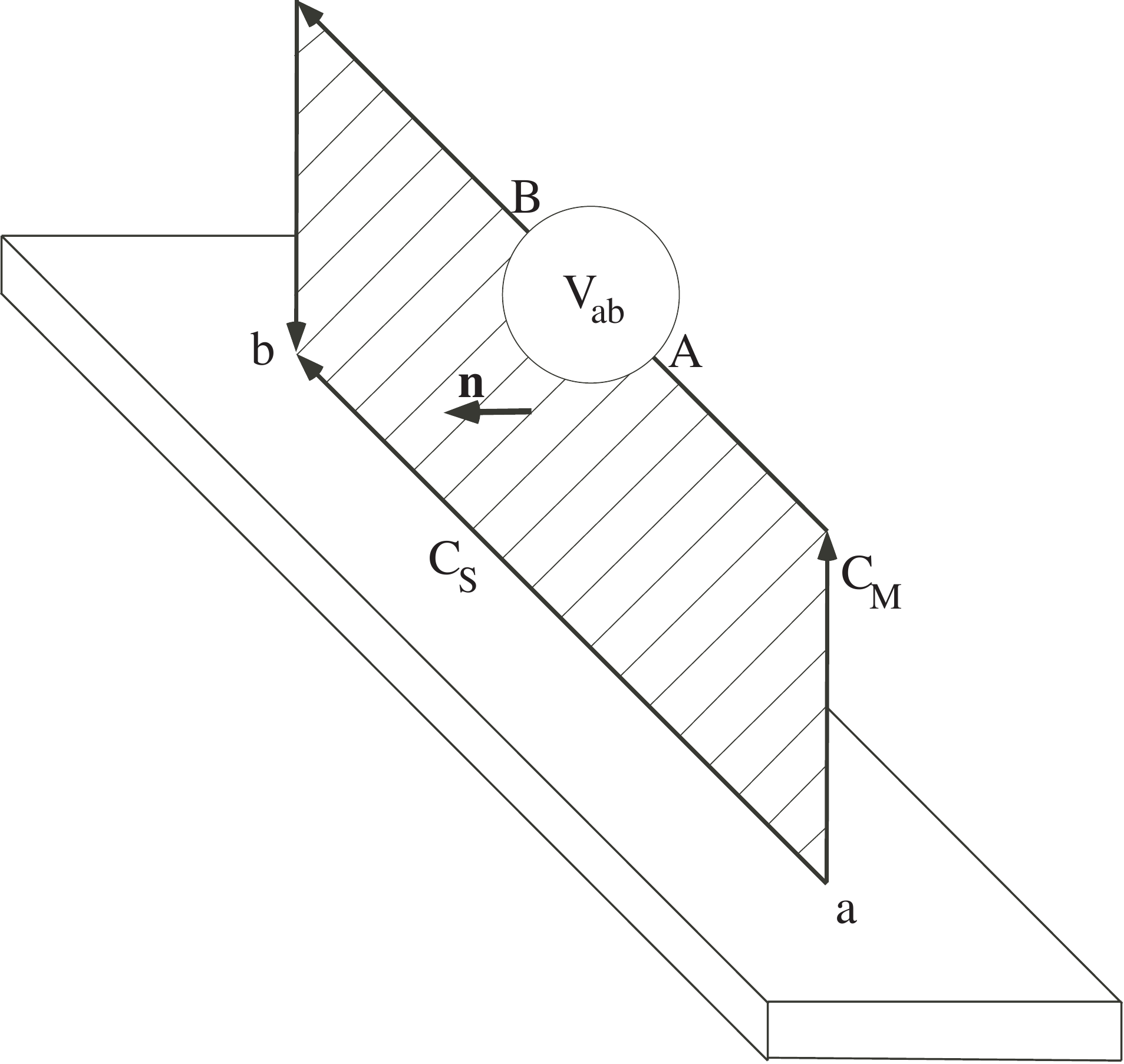}
\caption{%
Sketch of the measuring circuit used to determine the time-dependent voltage $V_{AB}=V_A-V_B$ across a high-impedance voltmeter when leads are attached to the sample at contacts $a$ and $b$. The surface $S_{MS}$ is shown cross-hatched. }
\label{VabSketch}
\end{figure}

Consider Fig.\ \ref{VabSketch}.  
According to  \cite{Clem70}, the voltage $V_{ab} = V_A-V_B$ across terminals $A$ and $B$ of a high-impedance voltmeter when low-resistance leads are connected to contacts $a$ and $b$ on a conductor is given by 
\begin{equation}
V_{ab}=\int_{a\;C_S}^b{\bm E} \cdot {\bm {dl}}-\frac{d\Phi_{MS}}{dt},
\label{Vab}
\end{equation}
where $C_S$ is any contour through the conductor from contact $a$ to contact $b$, $\bm E$ is the electric field in the conductor (the negative gradient of the electrochemical potential), 
\begin{equation}
\Phi_{MS} = \int _{S_{MS}}{\bm B} \cdot {\bm n} dS
\end{equation}
is the magnetic flux through the area $S_{MS}$ bounded by the sample contour $C_S$ and the path $C_M$ along the measuring circuit leads, $\bm B$ is the magnetic induction, and $\bm n$ is a unit vector perpendicular to the surface $S_{MS}$ as shown in Fig.\ 1.
As can be shown with the help of Faraday's law, $V_{ab}$ is independent of the chosen path $C_S$, because any change in the first term of equation (\ref{Vab}) is compensated for by a corresponding change in the second term.

\subsection{Helically wound cable}
Consider $N$ helically wound superconducting layers of radii $R_1 < R_2 < R_3 < ... < R_N$ with pitches $P_1$, $P_2$, ..., $P_N$. The closely spaced superconducting tapes  forming the helical layers are wound in either a right-handed or left-handed sense.  The resulting cable is centered on the $z$ axis. 
\begin{figure}%***** Fig.2 ************************
\includegraphics[width=6cm]{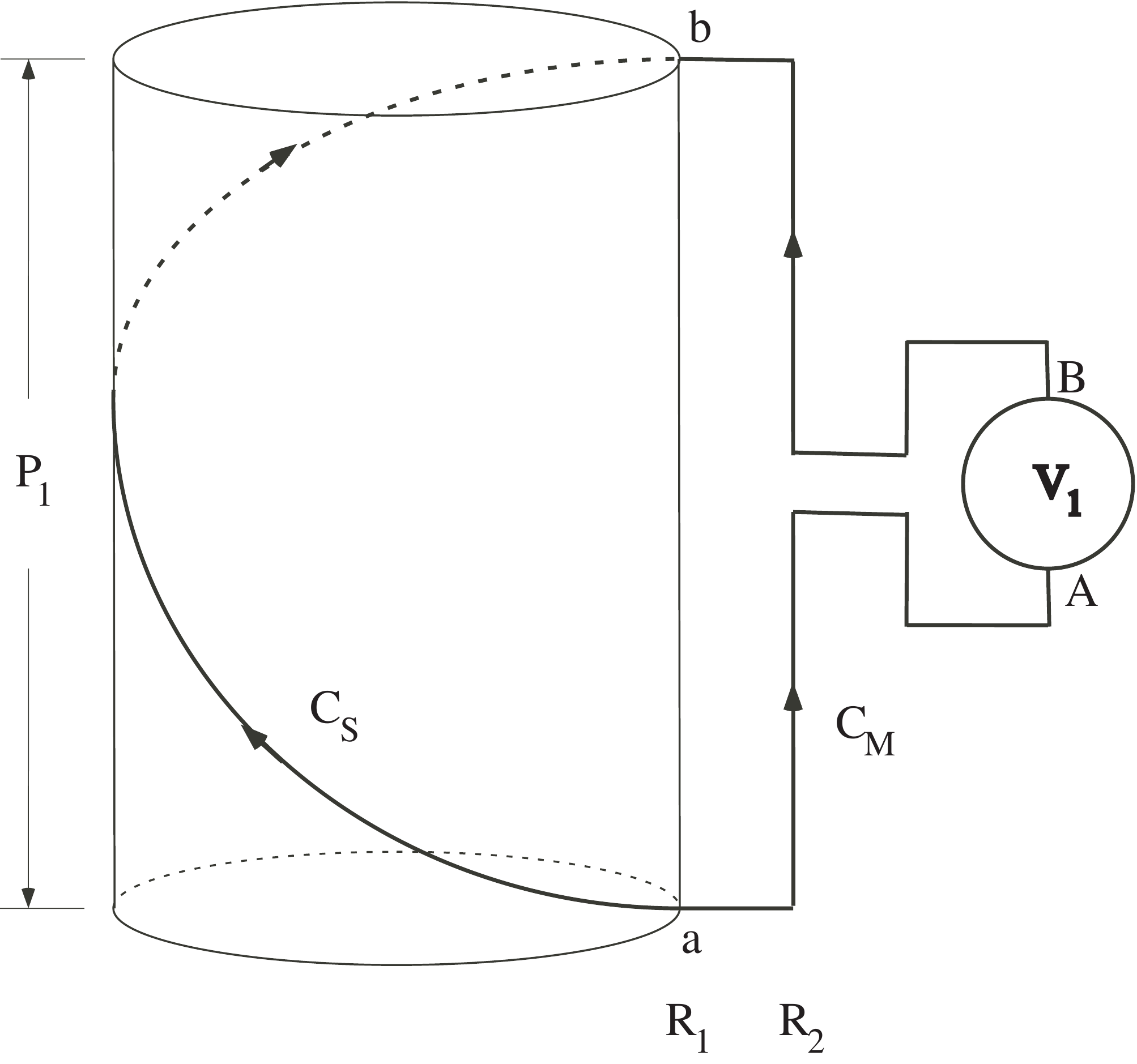}
\caption{%
Sketch of the measuring circuit used with equation (\ref{Vab}) to determine the measured voltage along the inner conductor (layer 1), consisting of tapes in the form of left-handed helices.  The spacing along the $z$ direction between the contacts $a$ and $b$ on one of the tapes is the pitch of layer 1, $P_1$.}  
\label{Layer1}
\end{figure}  
Figure \ref{Layer1} shows the geometry of the paths $C_S$ and $C_M$ for measuring the voltage $V_1$ along one pitch length $P_1$ of layer 1.  The measuring circuit leads are brought out radially from contacts $a$ and $b$, then laid along the outer surface of the cable in the $z$ direction, brought together, twisted (twist not shown), and then connected to terminals $A$ and $B$ of the voltmeter.  This figure shows the geometry for two layers, such that the outer surface of the cable is at radius $R_2$.  However, for $N$ layers, the outer surface of the cable is at radius $R_N$.

Using equation (\ref{Vab}) and figures like that of Fig.\ \ref{Layer1} for layers with left-handed pitch and figures with right-handed paths $C_S$  for the other layers, we obtain the following equations determining the voltage per unit length $V_n'$ for all $N$ layers when the current carried in the $z$ direction by layer $n$ is $I_n(t)$: 
\begin{equation}
V_n'=E_n+\frac{t_n}{2\pi R_n}\int_{\rho < R_n}\frac{dB_z}{dt}dS
+\int_{R_n}^{R_N}\frac{dB_\phi}{dt}d\rho,
\label{Vn'1}
\end{equation}
where
\begin{equation}
t_n = \tan \theta_n = \pm \frac{2\pi R_n}{P_n}
\end{equation}
is the tangent of the pitch angle of layer $n$,
with the positive (negative) sign for a right-handed (left-handed) helix, and where $E_n$ is the electric field along the length of a tape in layer $n$.   The first integral is over a surface of radius $R_n$ perpendicular to the $z$ axis, and the second integral is a line integral over the radial coordinate from $R_n$ to $R_N$.  Equation (\ref{Vn'1}) does not include the hysteretic critical-state penetration of magnetic flux into the superconducting layers.  When the thickness $d_s$ of a superconducting layer is much smaller than the spacing between layers, the effect of such flux penetration has a negligible effect upon the distribution of currents among the $N$ layers.  The ac losses associated with hysteresis within the superconducting layers therefore must be calculated separately. 

If the current  $I_n(t)$ in layer $n$ never exceeds the critical current $I_{cn}$ for that layer, the electric field $E_n$ can be neglected when determining how the total time-dependent current $I_t(t) =\sum_n I_n(t) = I \sin \omega t$ is distributed between the layers.  However, if the magnitude of the current in layer $n$ exceeds $I_{cn}$, it is important to account for the resulting  electric field $E_n$.  In such a case, one may write the resulting electric field as 
\begin{eqnarray}
E_n&=&\rho_f(\bar J_n - J_c), \; \bar J_n > J_c, 
\label{En1}\\
&=&0, \; |\bar J_n| \le J_c, \\
&=&\rho_f(\bar J_n + J_c), \; \bar J_n <- J_c,
\end{eqnarray}
where $\rho_f$ is the flux-flow resistivity, $\bar J_n$ is the current density along the tape direction in layer $n$ averaged across the superconductor's thickness $d_s$, and $J_c$ is the critical current density in the superconductor.
The electric field $E_n$ can be rewritten in terms of the current $I_n = 2 \pi R_n d_s \bar J_n \cos \theta_n$ carried by layer $n$ and its corresponding critical current $I_{cn}= 2 \pi R_n d_s J_c \cos \theta_n$ as follows:
\begin{eqnarray}
E_n&=&\rho_n(I_n - I_{cn}), \; I_n > I_{cn}, \\
&=&0, \;  |I_n| > I_{cn}, \\
&=&\rho_n( I_n + I_{cn},), \;  I_n >- I_{cn},
\label{En6}
\end{eqnarray}
where $\rho_n = \rho_f/ 2 \pi R_n d_s \cos \theta_n$.

It also is possible to model the electric field in the high-current range  using an expression of the form \cite{Daeumling99}
\begin{equation}
E_n = E_c(I_n/I_{cn})^m,
\label{powerlawE}
\end{equation} 
where $E_c$ is a very small electric field, $I_n >0$, and $m$ is a large power-law exponent, such as $m = 30$. However, here we use the forms of equations (\ref{En1})-(\ref{En6}). In this case the equations for $V_n'$ are linear first-order differential equations and all the $I_n$ are sinusoidal in time when (a) there is no hysteretic magnetic material between the superconducting layers and (b) $|I_n| < I_{cn}$ for all layers.
Mathematically, the properties of typical superconductors are such that the main effect of the terms involving $E_n$ is that whenever the magnitude of $I_n$ is driven to exceed $I_{cn}$, the value of $I_n$ is stabilized at approximately $\pm I_{cn}$, and the currents in the other layers adjust to satisfy this constraint.  This effect guarantees that the currents $I_{cn}(t)$ are not sinusoidal, even though the total current $I_t(t)$ is sinusoidal, and as a result the energy  loss per unit length associated with the nonlinearities can be calculated as the time integral over one cycle of the product of the current and the voltage per unit length.

To calculate the terms in equation (\ref{Vn'1}) involving time derivatives of integrals of the magnetic induction components $B_z$ and $B_\phi$, we first need to determine the magnetic field components $H_z$ and $H_\phi$, to which the magnetic induction responds.  These are obtained from Ampere's law as follows:
\begin{eqnarray}
H_\phi(\rho)&=&0, \; \rho<R_1, 
\label{Hphi1}\\
&=&\sum_{m=1}^n\frac{I_m}{2\pi \rho}, \;R_n <\rho < R_{n+1}, \; n\le N-1,
\label{Hphi2}\\
&=&\sum_{m=1}^N\frac{I_m}{2\pi \rho}, \;R_N <\rho, 
\label{Hphi3} \\
H_z(\rho)
&=&\sum_{m=n}^N\frac{I_m t_m}{2\pi R_m}, \;R_{n-1} <\rho < R_{n},
\label{Hz1} \\
&=&0, \; R_N <\rho,
\label{Hz2}
\end{eqnarray}
where in equation (\ref{Hz1}) we define $R_0 = 0$.

\subsection{Nonmagnetic substrates}

If all the components of the superconducting tapes are nonmagnetic, then the magnetic induction between tapes obeys ${\bm B} = \mu_0 {\bm H}$.  Evaluating the integrals in  equation (\ref{Vn'1}) then yields the measured voltage per unit length of cable,
\begin{equation}
V_n' = E_n + \sum_{m=1}^N L_{nm}'dI_m/dt,\label{Vn'}
\end{equation}
with the inductances per unit length 
\begin{equation}
L_{nm}'=\frac{\mu_0}{4\pi}\Big[\ln\Big(\frac{R_N^2}{R_{nm}^{>2}}\Big)+
\frac{R_{nm}^{<2}}{R_nR_m}t_n t_m\Big],
\label{Lnm'}
\end{equation}
where $R_{nm}^{>}$ ($R_{nm}^{<}$) is the larger (smaller) of $R_n$ and $R_m$ \cite{Garber76,Daeumling99,foot1}.
Equations (\ref{Vn'}) and (\ref{Lnm'}) are useful in calculating the impedance of this kind of superconducting cable in the grid.  

For long cables, inductive effects dominate over the electrical resistance at the ends, and we must have 
\begin{equation}
V_1' = V_2'= V_3'= ... = V_N',
\label{VnEquality}
\end{equation}
which yields $N-1$ equations relating the currents $I_n$.  If $|I_n| < I_{cn}$ for all layers, all the voltages $V_n'$ are sinusoidal.  Subtracting the equation for $V_N'$ from that for $V_n'$ and noting that $I_n(t) = I_{n0}\sin \omega t$, we obtain
\begin{equation}
\sum_{m=1}^N (L_{nm}'-L_{Nm}')I_{m0} = 0, \; n = 1, 2, ..., N-1.
\end{equation}
We can put this into dimensionless form by dividing the inductances by $\mu_0/4\pi$ and by dividing the $I_{m0}$ by $I$, the peak value of the total current, which yields
\begin{equation}
\sum_{m=1}^N \alpha_{nm}i_m = 0, \; n = 1, 2, ..., N-1,
\end{equation}
where 
\begin{equation}
\alpha_{nm}=\ln\Big(\frac{R_N^2}{R_{nm}^{>2}}\Big)+
\frac{R_{nm}^{<2}}{R_nR_m}t_n t_m-\frac{R_m}{R_N}t_Nt_m
\end{equation}
and
\begin{equation}
i_m = I_{m0}/I.
\end{equation}
Since the sum of the $i_m$ is 1 by definition and  $\alpha_{Nm}=1$,  we now have $N$ equations and $N$ unknowns:
\begin{equation}
\sum_{m=1}^N \alpha_{nm}i_m = \delta_{Nm}, \; n = 1, 2, ..., N.
\end{equation}

For the simplest case of $N = 2$ we have, for example,
\begin{eqnarray}
\alpha_{11} &=& \ln(R_2^2/R_1^2) + t_1^2-(R_1/R_2)t_1t_2, \\
\alpha_{12}&=&(R_1/R_2)t_1t_2 - t_2^2, \\
\alpha_{21}&=&1, \\
\alpha_{22}&=&1,
\end{eqnarray}
from which we obtain
\begin{eqnarray}
i_1&=&-\frac{\alpha_{12}}{\alpha_{11}-\alpha_{12}}, \\
i_2&=&\frac{\alpha_{11}}{\alpha_{11}-\alpha_{12}}.
\end{eqnarray}
Thus the ratio of the current amplitudes is
\begin{equation}
\frac{i_1}{i_2}=\frac{t_2^2-(R_1/R_2)t_1t_2}{\ln(R_2^2/R_1^2)+t_1^2-(R_1/R_2)t_1t_2}.
\label{i1byi2}
\end{equation}

If we wish to have $i_1= i_2$, there are four solutions, with $t_1 = \tan \theta_1$ positive or negative and $t_2 = \tan \theta_2$ positive or negative, so long as 
\begin{equation}
\tan^2\theta_2 = \tan^2\theta_1 +\ln(R_2^2/R_1^2).
\label{tantheta2}
\end{equation}

\section{Two superconducting layers\label{2layers}}

\subsection{Basic equations for magnetic substrates}

Consider next the influence of substrate layers separating the superconducting layers and the possibility that some of these substrate layers are magnetic.  Let substrate layer $n$ denote a substrate (or spacing layer) that separates superconducting layers of radii $R_{n-1}$ and $R_n$, where $R_0 = 0$.  
Let us denote the components of the magnetic field $\bm H$ between superconducting layers of radii $R_{n-1}$ and $R_n$ as  $H_{n\phi}$ and $H_{nz}$, such that from equations (\ref{Hphi1})-(\ref{Hz1})
\begin{eqnarray}
H_{1\phi}&=&0,
\label{H1phi}\\
H_{n\phi}&=&\sum_{m=1}^{n-1}\frac{I_m}{2\pi \rho}, \;n\ge 2,
\label{Hnphi}\\
H_{nz}
&=&\sum_{m=n}^N\frac{I_m t_m}{2\pi R_m}.
\label{Hnz}
\end{eqnarray}

Although it would not be difficult to generalize our results for an arbitrary number of superconducting layers and substrate layers, our main interest here is the case of two superconducting layers, for which the components of the magnetic field $\bm H$ are, from equations (\ref{H1phi})-(\ref{Hnz}),
\begin{eqnarray}
H_{1\phi}&=&0,
\label{H1phi2}\\
H_{2\phi}&=&\frac{I_1}{2\pi \rho},
\label{H2phi2}\\
H_{1z}
&=&\frac{I_1 t_1}{2\pi R_1}+\frac{I_2 t_2}{2\pi R_2}
\label{H1z2} \\
H_{2z}
&=&\frac{I_2 t_2}{2\pi R_2}.
\label{H2z2}
\end{eqnarray}

The voltage equations, from equation (\ref{Vn'1}), become
\begin{eqnarray}
V_1'&=& E_1 + \frac{t_1}{2\pi R_1} \dot{\Phi}_{1z}+\dot{\Phi}_{2\phi}',
\label{V1'} \\
V_2'&=& E_2 + \frac{t_2}{2\pi R_2} (\dot{\Phi}_{1z}+\dot{\Phi}_{2z}),
\label{V2'}
\end{eqnarray}
where the dots represent time derivatives of the magnetic flux contributions,
\begin{eqnarray}
\Phi_{1z}&=&\int_0^{R_1}2 \pi \rho B_{1z}(\rho)d\rho,
\label{Phi1z} \\
\Phi_{2z}&=&\int_{R_1}^{R_2}2 \pi \rho B_{2z}(\rho)d\rho,
\label{Phi2z} \\
\Phi_{2\phi}' &=&\int_{R_1}^{R_2}B_{2\phi}(\rho)d\rho. 
\label{Phi2phi}
\end{eqnarray}

Other than in the superconducting layers, we assume that the magnetic flux density can always be written as $\bm B(\rho) = \mu_0[\bm H(\rho) + \bm M(\rho)],$ where $\bm M(\rho) = 0$ in empty space or in a nonmagnetic substrate, $\bm M(\rho) = \chi \bm H(\rho)$ in a substrate that is paramagnetic with relative permeability $\mu = 1 + \chi$, or $\bm  M(\rho) = M_{FM}(\bm H)\hat H$ in a ferromagnetic substrate, where its magnetization is given by a hysteresis function $M_{FM}(\bm H)$ that depends upon  whether $\bm H$ is increasing or decreasing, and $\hat H = \bm H/|\bm H|$.  Accordingly, each of the magnetic flux contributions given in equations (\ref{Phi1z})-(\ref{Phi2phi}) can be separated into two terms, one due to $\bm H$ alone and the other due to $\bm M$:
\begin{eqnarray}
\Phi_{1z}&=&\Phi_{1Hz}+\Phi_{1Mz},\\
\Phi_{2z}&=&\Phi_{2Hz}+\Phi_{2Mz},\\
\Phi_{2\phi}' &=&\Phi_{2H\phi}'+\Phi_{2M\phi}',
\end{eqnarray}
where
\begin{eqnarray}
\Phi_{1Hz}&=&\mu_0\int_0^{R_1}2 \pi \rho H_{1z}d\rho,
\label{Phi1Hz}\\
\Phi_{2Hz}&=&\mu_0\int_{R_1}^{R_2}2 \pi \rho H_{2z}d\rho,
\label{Phi2Hz}\\
\Phi_{2H\phi}' &=&\mu_0\int_{R_1}^{R_2}H_{2\phi}(\rho)d\rho,
\label{Phi2Hphi'}\\ 
\Phi_{1Mz}&=&\mu_0\int_0^{R_1}2 \pi \rho M_{1z}(\rho)d\rho,
\label{Phi1HMz}\\
\Phi_{2Mz}&=&\mu_0\int_{R_1}^{R_2}2 \pi \rho M_{2z}(\rho)d\rho,
\label{Phi2Mz}\\
\Phi_{2M\phi}' &=&\mu_0\int_{R_1}^{R_2}M_{2\phi}(\rho)d\rho.
\label{Phi2Mphi'} 
\end{eqnarray}

Equations (\ref{V1'}) and (\ref{V2'}) become
\begin{eqnarray}
V_1'\!&\!=\!& \!E_1 \!+ \!\frac{t_1}{2\pi R_1} (\dot{\Phi}_{1Hz}\!+\!\dot{\Phi}_{1Mz})\!+\!(\dot{\Phi}_{2H\phi}'\!+\!\dot{\Phi}_{2M\phi}'),
\label{V1'HM}\\
V_2'\!&\!=\!& \!E_2 \!+\! \frac{t_2}{2\pi R_2} (\dot{\Phi}_{1Hz}\!+\!\dot{\Phi}_{1Mz}\!+\!\dot{\Phi}_{2Hz}\!+\!\dot{\Phi}_{2Mz}).
\label{V2'HM}
\end{eqnarray}
where the dots again represent time derivatives of the magnetic flux contributions.
For long cables, inductive effects dominate over the electrical resistance at the ends, and we must have 
\begin{equation}
V_1' = V_2'.
\label{V1'V2'Equality}
\end{equation}

\subsection{Paramagnetic substrates}

The complex behavior when the substrate layers are ferromagnetic is beyond the scope of the present paper.  Here we focus on the case when one or two of the substrate layers are paramagnetic.  
 Let $p_n$ be a factor that  tells whether substrate layer $n$ is paramagnetic or not.  That is, if $p_n = 1$, this will indicate that substrate layer $n$ is paramagnetic with relative permeability $\mu = 1+\chi$, where $\chi$ is independent of the magnetic field.  Of primary interest here is the case for which $\chi \gg 1$, but we will assume arbitrary values of $\chi$.  If $p_n = 0$, this will indicate that substrate layer $n$ is nonmagnetic, in which case its relative permeability is $\mu = 1$ and its magnetic susceptibility is $\chi = 0$.  We denote the thickness of a substrate layer $n$ as $d_n$, where $d_n \ll R_n$.  Substrate layer 1 is the layer on which superconducting layer 1 rests and thus has a radius very nearly equal to $R_1$.   

Carrying out the integrals in equations (\ref{Phi1Hz})-(\ref{Phi2Mphi'}), we obtain
\begin{eqnarray}
\Phi_{1Hz}&=&\mu_0\pi R_1^2H_{1z},
\label{Phi1Hzp}\\
\Phi_{2Hz}&=&\mu_0 \pi (R_2^2-R_1^2) H_{2z},
\label{Phi2Hzp}\\
\Phi_{2H\phi}' &=&\frac{\mu_0 I_1}{2 \pi}\ln(R_2/R_1),
\label{Phi2Hphi'p}\\ 
\Phi_{1Mz}&=&p_1\mu_0 2 \pi R_1 d_1 H_{1z} \chi,
\label{Phi1HMzp}\\
\Phi_{2Mz}&=&p_2\mu_0 2 \pi R_2 d_2 H_{2z} \chi,
\label{Phi2Mzp}\\
\Phi_{2M\phi}' &=&p_2\frac{\mu_0 I_1}{2 \pi R_2}d_2 \chi.
\label{Phi2Mphi'p} 
\end{eqnarray}
Substituting these expressions into equations (\ref{V1'HM}) and (\ref{V2'HM}), making use of equations (\ref{H1z2}) and (\ref{H2z2}), we obtain
\begin{eqnarray}
V_1'&= &E_1 +L_{11}'{\dot I}_1+L_{12}'{\dot I}_2, 
\label{V1'p}\\
V_2'&=&E_2 +L_{21}'{\dot I}_1+L_{22}'{\dot I}_2, 
\label{V2'p}
\end{eqnarray}
where
\begin{eqnarray}
L_{11}' &=&\frac{\mu_0}{4\pi}[t_1^2(1+2p_1d_1\chi/R_1)\nonumber \\
&&+\ln(R_2^2/R_1^2)+2p_2d_2\chi/R_2],
\label{L11p}\\
L_{12}'&=&L_{21}'= 
\frac{\mu_0}{4\pi}(R_1/R_2)t_1t_2(1+2p_1d_1\chi/R_1),
\label{L12p}\\
L_{22}'&=&\frac{\mu_0}{4\pi}t_2^2(1+2p_1R_1d_1\chi/R_2^2+2p_2d_2\chi/R_2).
\label{L22p}
\end{eqnarray}

Recall that $V_1'=V_2'$.  If the magnitudes of neither $I_1$ nor $I_2$ exceeds its critical current, then $E_1=E_2 = 0$, and we find from equations (\ref{V1'p}) and (\ref{V2'p}) that ${\dot I}_1$ and ${\dot I}_2$ obey 
\begin{equation}
{\dot I}_1= r_{12} {\dot I_2},
\label{I1dot}
\end{equation}
where
\begin{equation}
r_{12} = \frac{L_{22}'-L_{12}'}{L_{11}'-L_{21}'}.
\label{r12}
\end{equation}
Thus $I_1=r_{12}I_2$+const.  However, since the total current is  $I_1(t) + I_2(t)=I_t(t) = I \sin \omega t$, we must have const = 0, such that $I_1 = I_{10}\sin \omega t$ and $I_2 = I_{20}\sin \omega t$.  Defining $i_1 = I_{10}/I$ and $i_2=I_{20}/I$ as the fraction of the total current carried by layers 1 and 2, we obtain 
\begin{equation}
\frac{i_1}{i_2}=r_{12}=\frac{L_{22}'-L_{12}'}{L_{11}'-L_{21}'}.
\label{i1byi2p}
\end{equation}
Note that this equation reduces to equation (\ref{i1byi2}) when $\chi = 0$.

For the general {\it in-out} ($p_1=1$ and $p_2 = 0$) case with paramagnetic substrates,  when $\chi \gg 1$, the inductances are dominated by the large terms in equations (\ref{L11p})-(\ref{L22p}) proportional to $\chi$, and $r_{12}$ in 
equation (\ref{r12}) then can be expanded to first order in $1/\chi$ as 
\begin{equation}
r_{12} = -\frac{R_1 t_2}{R_2 t_1}+\frac{R_1 t_2 [R_1 R_2 \ln (R_2^2/R_1^2)+ (R_2^2-R_1^2)t_1 t_2]}{2 d_1R_2 t_1^2 (R_2 t_1 -R_1 t_2) \chi}.
\end{equation}
Using $I_1(t) + I_2(t)=I_t(t) = I \sin \omega t$ and  equation (\ref{H1z2}), we then obtain, to first order in $1/\chi$,
\begin{equation}
H_{1z}=\frac{t_2 [R_1 R_2 \ln (R_2^2/R_1^2)+ (R_2^2-R_1^2)t_1 t_2]}{4 \pi d_1 ( R_2 t_1 -R_1 t_2)^2 \chi}I_t.
\label{H1zPMinout}
\end{equation}
Thus, because of the magnetic flux in the inner paramagnetic substrate, the currents induced in the two superconducting layers drive the axial magnetic field to very small values, and even to zero in the limit as $\chi \to \infty$. Although the magnetization obeys 
$M_{1z} = H_{1z} \chi$, $H_{1z}$ is proportional to $1/\chi$, such that 
\begin{equation}
M_{1z} = \frac{t_2 [R_1 R_2 \ln (R_2^2/R_1^2)+ (R_2^2-R_1^2)t_1 t_2]}{4 \pi d_1  ( R_2 t_1 -R_1 t_2)^2}I_t
\label{Mpinout}
\end{equation} 
remains finite.  

For the general {\it out-in} ($p_1=0$ and $p_2 = 1$) case with paramagnetic substrates,  when $\chi \gg 1$, the inductances again are dominated by the large terms in equations (\ref{L11p})-(\ref{L22p}) proportional to $\chi$, and $r_{12}$ in 
equation (\ref{r12}) again can be expanded to first order in $1/\chi$ as 
\begin{equation}
r_{12} = t_2^2 +\frac{t_2 \{R_2t_2[1-t_1^2-\ln (R_2^2/R_1^2)]-R_1t_1(1-t_2^2\}}{2 d_2 \chi}.
\end{equation}
In contrast to the {\it in-out} case, the leading terms in the expansion of the magnetic field $H_2=\sqrt{H_{2\phi}^2+H_{2z}^2}$ do not cancel, and 
using equations (\ref{H1phi2}), (\ref{H1z2}), and $I_1(t) + I_2(t)=I_t(t) = I \sin \omega t$ we then obtain
\begin{equation}
H_2 \approx \frac{t_2} {2 \pi  R_2 \sqrt{1+t_2^2}}I_t
\end{equation}
in the limit of large $\chi$.  Since the magnetization of the total substrate thickness $d_2$ between the two superconducting layers obeys 
$M_{2} = H_{2} \chi$, we have 
\begin{equation}
M_2 \approx  \frac{t_2 \chi} {2 \pi  R_2 \sqrt{1+t_2^2}}I_t
\label{Mpoutin}
\end{equation} 
in the limit of large $\chi$. 

\subsubsection{Cable with a single layer winding using a double-layer wire}

Consider a single-layer helical cable consisting of 20 tapes of width 4.4 mm with each tape having two ``inserts" of superconductor and substrate solder-bonded together at their superconductor faces.  This configuration is sometimes called ``face-to-face".  Such tape has double the critical current of a tape with a single insert and has been successfully fabricated at AMSC \cite{Rupich04}.  To see what impact this configuration has on ac loss, we can use the above formalism by choosing $\theta_2 = 20.58^\circ$ and  $\theta_1 = 19.42^\circ$ to give an average pitch angle of $20^\circ$.  The superconducting layers are assumed to have thickness 1 $\mu$m and to be separated by a spacing layer of 15 $\mu$m.  Let us choose the average radius of the outer superconducting layer to  be $R_2 = 15.000$ mm,  such that the average radius of the inner superconducting layer is $R_1 = 14. 084$ mm.  The pitch is $P = P_1 = P_2 = 251.04$ mm.  Note that since $t_1 = \tan \theta_1 = 2\pi R_2/P$ and $t_2 = \tan \theta_2/P$,  in equations (\ref{V1'p}) and   (\ref{V2'p}) we have $t_1/2\pi R_1 = t_2/2\pi R_2 = 1/P$.  Let us assume a paramagnetic substrate of thickness $d_1$ = 75 $\mu$m; so $p_1$ = 1, and we also take the spacing layer (solder) to be nonmagnetic; so $p_2$ = 0.
Equation (\ref{i1byi2p}) yields, {\it for all values of $\chi$}, the current ratio $i_1/i_2$ = 0.132, which means that $i_1$ = 0.117 and $i_2$ = 0.883.  However, since $I_{c1} = I_{c2} = I_c/2$, this means that the sinusoidal behavior of $I_1 = I_{10}\sin \omega t$ and $I_2 = I_{20}\sin \omega t$ holds only so long as the total current amplitude obeys $F = I/I_c < 1/2i_2 = 0.566= F_{ft}$.  

For increasing currents for which $F>F_{ft}$, the threshold value for {\it flux transfer}, the current in layer 2 is being driven above its critical current, and $E_2 > 0$.  However, the current $I_2$ in layer 2  will remain close to $I_{c2} = I_c/2$ while magnetic flux transfers into the spacing layer, allowing the excess current to transfer into layer 1.  When the current $I_t(t)$ reaches its maximum value, flux transfer stops, the amount of magnetic flux in the spacing layer remains frozen for a while, and the currents in the two layers again become sinusoidal in time, but not exactly in phase with the applied current $I_t(t)$.  We shall give a detailed calculation of this phenomenon in section 4 below.

It is important to note here that, although the above calculation was done assuming a paramagnetic substrate inside superconducting layer 1, the conclusion that the current distributions between layers 1 and 2 are unaffected by the magnetism holds even if the substrate is ferromagnetic.  The reason for this is that the terms involving ${\dot \Phi}_{1Mz}$, which describe the influence of the magnetic substrate inside $R_1$, cancel when $V_1'$ is set equal to $V_2'$ because of the  relation $t_1/2\pi R_1 = t_2/2\pi R_2 = 1/P$ , which holds for the face-to-face configuration. 
The conclusion is that double-layer face-to-face wire is not favorable for a single layer helical ac cable winding because, unless the current is kept very low (i. e., rms current below $0.566/\sqrt{2} = 40\%$ of the cable $I_c$), high ac flux-transfer losses will arise.  Of course, for DC applications, such face-to-face wire can be attractive because of its very high $I_c$.  

\subsubsection{Two-layer counter-wound cable}

Consider what happens when we have a double-layer helical cable consisting of $N$ = 20 tapes of width $w$ = 4.4 mm.  Let us assume that the inner superconducting layer has a radius of 15.0 mm and is wound as a left-handed helix with a pitch angle of $\theta_1$ = -20.97$^\circ$, such that $2\pi R_1 \cos\theta_1 = Nw = 88$ mm.  The outer superconducting layer has a radius of 15.5 mm and is wound as a right-handed helix with a pitch angle of $\theta_2$ = 25.37$^\circ$, such that $2\pi R_2 \cos\theta_2 = Nw = 88$ mm.  For $\chi= 0$, equation (\ref{i1byi2p}) yields $i_1/i_2 = 1.031.$  Now we consider the effect of the orientations of paramagnetic substrates, as an indication of what one might expect from the actual ferromagnetic substrates used in some wires \cite{Rupich12}.

{\it in-in}

For the {\it in-in} configuration,   assume that $p_1 = 1$ and there is a paramagnetic layer of thickness $d_1$ = 75 $\mu$m just inside the radius $R_1$. Also   assume that $p_2 = 1$ and there is a paramagnetic layer of thickness $d_2$ = 75 $\mu$m just inside the radius $R_2$.

{\it in-out}

For the {\it in-out}  configuration,   assume that $p_1 = 1$ and there is a paramagnetic layer of thickness $d_1$ = 75 $\mu$m just inside the radius $R_1$.  Assume that $p_2 = 0$ and there is  no paramagnetic layer just inside the radius $R_2$. 

{\it out-in}

For the {\it out-in} configuration,   assume that $p_1 = 0$ and there is no paramagnetic layer  just inside the radius $R_1$.  For simplicity,    assume that $p_2 = 1$ and there is a single paramagnetic layer of thickness $d_2$ = 150 $\mu$m just inside the radius $R_2$.  

{\it out-out}

For the {\it out-out} configuration,   assume that $p_1 = 0$ and there is no paramagnetic layer just inside the radius $R_1$.   For simplicity,   assume that $p_2 = 1$ and there is a paramagnetic layer of thickness $d_2$ = 75 $\mu$m just inside the radius $R_2$. 

Magnetic layers outside the radius $R_2$ cannot affect how the total current distributes between the two layers.

\begin{figure}%***** Fig.3 ************************
\includegraphics[width=8cm]{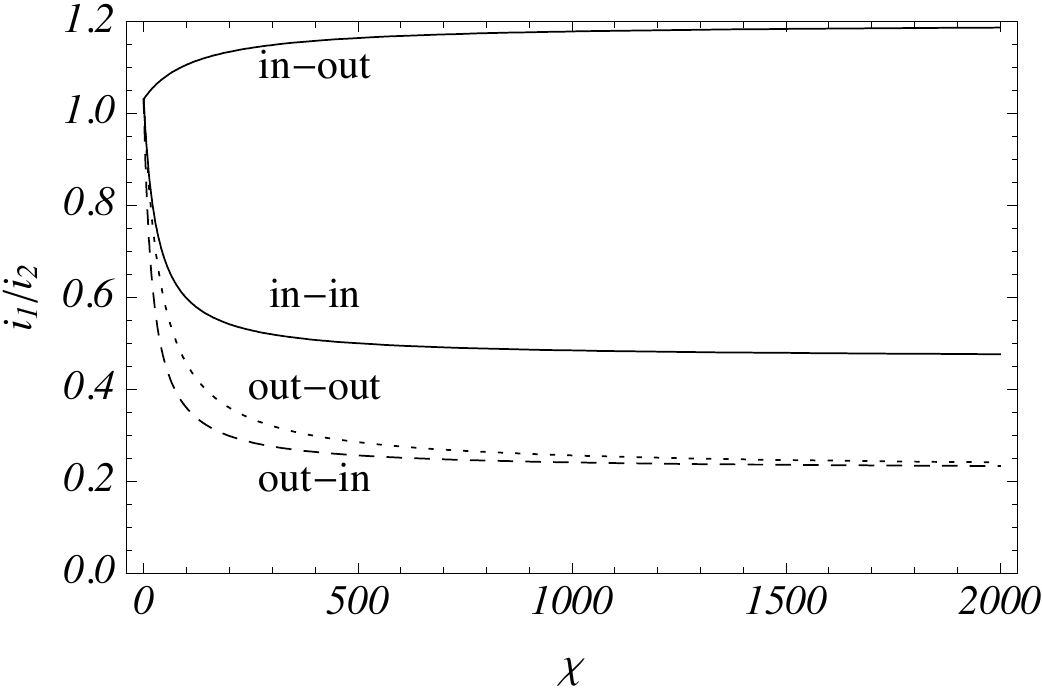}
\caption{%
Current ratio $i_1/i_2$ vs $\chi$ assuming paramagnetic substrates and subcritical current amplitudes $I_{10}< I_{c1}= I_c/2$ and $I_{20}< I_{c2}=I_c/2$ for the four cases discussed in the text: {\it in-in, in-out, out-in}, and {\it out-out}.}  
\label{344cable}
\end{figure}

As previously shown in [7], figure \ref{344cable} shows numerical results for $i_1/i_2$ vs $\chi$  obtained from equation (\ref{i1byi2p}), showing how the current ratio is  affected by paramagnetism in the substrates for the four cases.  Although these results do not apply to hysteretic ferromagnetic substrates, they provide some hints as to what  we might expect.  The most dramatic effects upon the current ratio $i_1/i_2$ occur when there is a magnetic substrate confined between the two superconducting layers.  This suggests that, with ferromagnetic substrates, switching effects might be most severe when at least one of these substrates lies between the two superconducting layers.  For the {\it in-out} configuration, there is no paramagnetic material between the two superconducting layers, and the current ratio changes only from 1.03 to 1.19 as $\chi$ increases from 0 to 2000.  This suggests that ferromagnetic switching effects might be minimized using this configuration. 

\section{AC Flux-transfer losses\label{fluxtransfer}}

We now consider the case of helically wound cables consisting of two thin superconducting layers with zero, one, or two paramagnetic substrates as in section \ref{2layers}.  Since the formalism of section \ref{2layers} neglects the hysteretic penetration of magnetic flux into the thin superconducting layers, one cannot calculate the ac losses due to the superconducting hysteresis by doing a time integral over one period of the voltage times the current.  In other words, the small superconducting hysteretic ac losses in the superconducting layers must be calculated separately. If we deal with paramagnetic substrates, these do not contribute, assuming we can neglect eddy-current losses. What we wish to address here are the ac flux-transfer losses that occur when the amplitude $I$ of the total current applied to the cable is large enough that the current in one of the two superconducting layers is driven above its critical current.  We denote the total current amplitude at which this occurs as $I_{ft}$, where the subscripts remind us that flux transfer occurs when $I > I_{ft}$.  If we use Norris symbol \cite{Norris70} $F = I/I_c$, then $F_{ft} = I_{ft}/I_c$.  We wish to calculate the large excess losses that occur when $F_{ft} < F < 1$.  

First we note that when $I < I_{ft}$, we have $I_{10} < I_{c1}$ and $I_{20} < I_{c2}$.  For this range of total current amplitudes, we have, from equation (\ref{i1byi2p}),
\begin{equation}
\frac{I_{10}}{I_{20}} = r_{12}=\frac{i_1}{i_2}=\frac{L_{22}'-L_{12}'}{L_{11}'-L_{21}'},
\label{I10byI20p}
\end{equation}
such that 
\begin{eqnarray}
I_{10}&=&\frac{r_{12}}{1+r_{12}}I, \\
I_{20}&=&\frac{1}{1+r_{12}}I.
\end{eqnarray}

Let us introduce a quantity similar to $r_{12}$ but instead giving the ratio of the critical currents of layers 1 and 2:
\begin{equation}
r_c = I_{c1}/I_{c2}
\label{rc}
\end{equation}

The details of what happens when $I > I_{ft}$ depend upon the ratio of $r_{12}/r_c$.  Consider first the case for which $r_{12}<r_c$.  In this case, we see that as the current amplitude $I$ increases, layer 2  reaches its critical current first.  This occurs when $I = I_{ft}$, which for $r_{12}<r_c$ is determined by the equation
\begin{equation}
I_{20}=\frac{1}{1+r_{12}}I_{ft}=I_{c2},
\end{equation}
such that 
\begin{equation}
F_{ft} = \frac{I_{ft}}{I_c}= \frac{(1+r_{12})I_{c2}}{I_c}.
\label{Fft}
\end{equation}
Note that when $I = I_{ft}$, 
\begin{equation}
I_{10}=\frac{r_{12}}{1+r_{12}}I_{ft}=r_{12}I_{c2}=
\frac{r_{12}}{r_c}I_{c1}< I_{c1}.
\end{equation}

Now consider what happens as the total applied current $I_t(t) = I \sin \omega t$ goes through one cycle.  According to critical-state concepts, if $I_t(t)$ increases with time and goes above $I_{ft}$, such that $I_2$ is driven slightly above $I_{c2}$, we still have $E_1 = 0$ in equation (\ref{V1'p}), but the electric field term $E_2$ in equation (\ref{V2'p}) becomes greater than zero.  The current $I_2(t)$ then stabilizes at the value $I_{c2}$ and the current in layer 1 becomes $I_1(t) = I_t(t) - I_{c2}$.  At $\omega t = \pi/2$, $I_t(t)$  reaches its maximum value $I$, and we have $I_1= I- I_{c2}$.  As $I_t(t)$ then decreases,  we again have $E_1 = E_2 = 0$, such that equation (\ref{I1dot}) yields ${\dot I}_1=r_{12} {\dot I}_2$ , whose solution is
\begin{eqnarray}
I_{1}(t)&=&(I-I_{c2})-\frac{r_{12}I(1-\sin \omega t)}{1+r_{12}},\\
I_{2}(t)&=&I_{c2}-\frac{I(1-\sin \omega t)}{1+r_{12}}. 
\end{eqnarray}
These equations hold for times obeying $\pi/2 \le \omega t \le \omega t_{min}$, where $t_{min}$ is the time that $I_2$ plateaus at its minimum value, 
\begin{equation}
I_{2}(t_{min})=I_{c2}-\frac{I(1-\sin \omega t_{min})}{1+r_{12}}=-I_{c2},
\end{equation}
such that 
\begin{equation}
\sin\omega t_{min} = 1-2F_{ft}/F,
\end{equation}
where $F_{ft}< F < 1$.
For times such that $\omega t_{min} \le \omega t \le 3\pi/2$, we have $I_2(t) = -I_{c2}$ and $I_1(t) = I_t(t) +I_{c2}$.  When $\omega t = 3\pi/2,$ we have $I_1 = -(I-I_{c2})$ and $I_2 = -I_{c2}$. 

During the next half cycle, for times obeying $3\pi/2 < \omega t < 5\pi/2$, or since the behavior is periodic with period $T = 2\pi/\omega$, for times obeying $-\pi/2 < \omega t < \pi/2$, we have the following behavior.  At time $\omega t = -\pi/2$ we have $I_1 = -(I-I_{c2})$ and $I_2 = -I_{c2}$.  As $I_t(t)$ increases, we again have $E_1 = E_2 = 0$, such that equation (\ref{I1dot})  yields ${\dot I}_1=r_{12} {\dot I}_2$, whose solution is now
\begin{eqnarray}
I_{1}(t)&=&-(I-I_{c2})+\frac{r_{12}I(1+\sin \omega t)}{1+r_{12}},\\
I_{2}(t)&=&-I_{c2}+\frac{I(1+\sin \omega t)}{1+r_{12}}. 
\end{eqnarray}
These equations hold  for times obeying $-\pi/2 \le \omega t \le \omega t_{max}$, where $t_{max}$ is the time that $I_2$ plateaus at its maximum value,  
\begin{equation}
I_{2}(t_{max})=-I_{c2}+\frac{I(1+\sin \omega t_{max})}{1+r_{12}}=I_{c2},
\end{equation}
such that 
\begin{equation}
\sin\omega t_{max} = 2F_{ft}/F-1,
\label{omegatmax}
\end{equation}
where $F_{ft}< F < 1$.
For times such that $\omega t_{max} \le \omega t \le \pi/2$, we have $I_2(t) = I_{c2}$ and $I_1(t) = I_t(t) -I_{c2}$.  
\begin{figure}%***** Fig.4 ************************
\includegraphics[width=10cm]{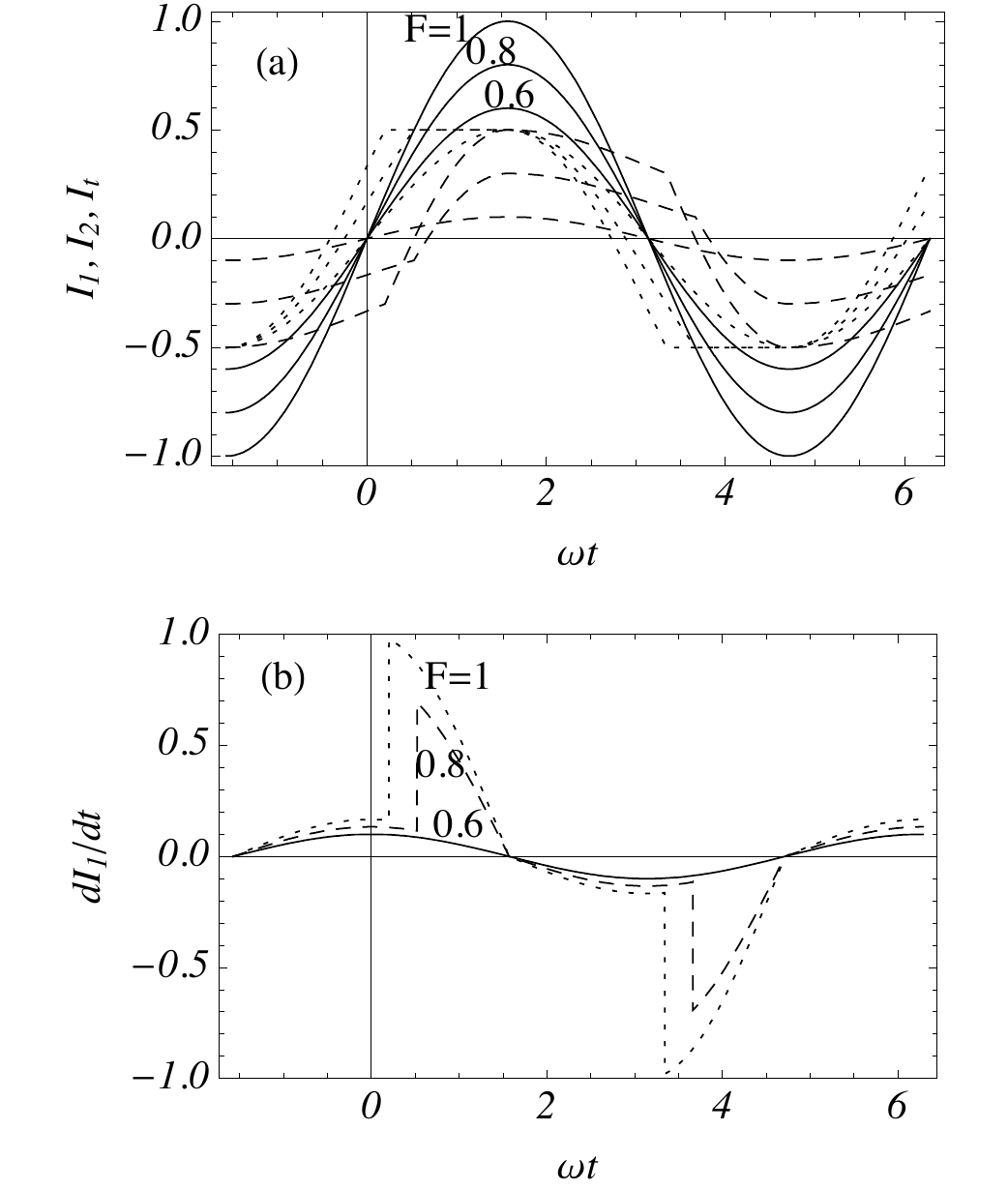}
\caption{%
(a) $I_1(t)$ (dashed), $I_2(t)$ (dotted), and $I_t(t) = I_1(t)+I_2(t) = I \sin \omega t$ (solid, all currents normalized to the total critical current $I_c$) vs $\omega t$ for $r_{12} = 0.2$, $r_c = 1$, $F_{ft} = 0.6$, and values of $F = I/I_c$ = 0.6, 0.8, and 1.  When $F < F_{ft}$,  $I_1(t)$ and  $I_2(t)$ are both sinusoidal in the ratio $I_1/I_2 = r_{12}$, and there are no flux-transfer losses.  However, when $F > F_{ft}$, $I_2(t)$ saturates at the values $\pm I_{c2} = \pm I_c/2$ over parts of the ac cycle, and $I_1(t)$ picks up the difference as magnetic flux transfers into the space between layers 1 and 2. (b) $dI_1/dt$ (normalized to $I_c \omega$)  vs $\omega t$ for $r_{12} = 0.2$, $F_{ft} = 0.6$, and values of $F$ = 0.6 (solid), 0.8 (dashed), and 1 (dotted). The flux-transfer loss per cycle per unit length, which  is proportional to the time integral over one cycle of $(dI_1/dt)I_t(t)$, is zero at $F = F_{ft}$ and is proportional to $F-F_{ft}$ for $F > F_{ft}$.}  
\label{4ab}
\end{figure}  

Shown in Fig.\ \ref{4ab}(a) are plots of $I_1(t),$ $I_2(t)$, and $I_t(t) = I_1(t)+I_2(t) = I \sin \omega t$ calculated from equations (\ref{Fft})-(\ref{omegatmax}). Fig.\ \ref{4ab}(b) shows plots of $dI_1/dt$ for various values of $F = I/I_c \ge F_{ft}$.  

The ac flux-transfer loss per cycle per unit length of cable is given by 
\begin{equation}
Q_{ft}' = \int_0^T V_1'(t)I_t(t) dt, 
\end{equation}
where $T = 2\pi/\omega$ is the period of one cycle.  Since $E_1 = 0$ at all times and $I_2 = I_t - I_1$, we have
\begin{equation}
V_1' = (L_{11}'-L_{12}')dI_1/dt + L_{12}'dI_t/dt.
\end{equation}
However, since the second term on the right-hand side is a purely inductive contribution, it does not contribute to the ac losses.  Thus $Q_{ft}'$ is proportional to the time integral over one period of the product of $dI_1/dt$ and $I_t(t) = I \sin \omega t.$  The result for $r_{12} < r_c$ can be expressed as 
\begin{equation}
Q_{ft}' = 4 (L_{11}'-L_{12}')I_c I_{c2}(F-F_{ft}),
\label{Qftr12small}
\end{equation}
where $F_{ft}$ is given by equation (\ref{Fft}).  Equation (\ref{Qftr12small}) also can be obtained by noting that energy is dissipated only over those portions of the cycle for which $E_2 \ne 0$, $|I_2| \approx I_{c2}$, and $E_2 = (L_{11}'-L_{21}'){\dot I}_1$ [see equations (\ref{V1'p}) and (\ref{V2'p})].  The coefficient of $I_{c2}$ on the right-hand side of equation (\ref{Qftr12small}) is twice the magnitude of the magnetic flux per unit length that passes through the superconducting layer 2 during the resistive portion of a half cycle.

Let us next examine what happens when $I > I_{ft}$ in the case for which $r_{12}>r_c$.  We see that in this case as the current amplitude $I$ increases, layer 1 will reach its critical current first.  This occurs when $I = I_{ft}$, which is now determined from the equation
\begin{equation}
I_{10}=\frac{r_{12}}{1+r_{12}}I_{ft}=I_{c1},
\end{equation}
such that 
\begin{equation}
F_{ft} = \frac{I_{ft}}{I_c}= \frac{(1+r_{12})I_{c1}}{r_{12}I_c}.
\label{Fft2nd}
\end{equation}
Note that when $I = I_{ft}$, 
\begin{equation}
I_{20}=\frac{1}{1+r_{12}}I_{ft}=\frac{I_{c1}}{r_{12}}=
\frac{r_c}{r_{12}}I_{c2}< I_{c2}.
\end{equation}

Now consider what happens as the total applied current $I_t(t) = I \sin \omega t$ goes through one cycle.  According to critical-state concepts, if $I_t(t)$ increases with time and goes above $I_{ft}$, such that $I_1$ is driven slightly above $I_{c1}$, we still have $E_2 = 0$ in equation (\ref{V2'p}), but the electric field term $E_1$ in equation (\ref{V1'p}) becomes greater than zero.  The current $I_1(t)$ then stabilizes at the value $I_{c1}$ and the current in layer 2 becomes $I_2(t) = I_t(t) - I_{c1}$.  At $\omega t = \pi/2$, $I_t(t)$  reaches its maximum value $I$, and we have $I_2= I- I_{c1}$.  As $I_t(t)$ then decreases,  we again have $E_1 = E_2 = 0$, such that equation (\ref{I1dot}) yields ${\dot I}_1=r_{12} {\dot I}_2$ , whose solution is
\begin{eqnarray}
I_{1}(t)&=&I_{c1}-\frac{r_{12}I(1-\sin \omega t)}{1+r_{12}},\\
I_{2}(t)&=&(I-I_{c1})-\frac{I(1-\sin \omega t)}{1+r_{12}}.
\end{eqnarray}
These equations hold for times obeying $\pi/2 \le \omega t \le \omega t_{min}$, where $t_{min}$ is the time that $I_1$ plateaus at its minimum value, 
\begin{equation}
I_{1}(t_{min})=I_{c1}-\frac{r_{12}I(1-\sin \omega t_{min})}{1+r_{12}}=-I_{c1},
\end{equation}
such that 
\begin{equation}
\sin\omega t_{min} = 1-2F_{ft}/F,
\end{equation}
where $F_{ft}< F < 1$.
For times such that $\omega t_{min} \le \omega t \le 3\pi/2$, we have $I_1(t) = -I_{c1}$ and $I_2(t) = I_t(t) +I_{c1}$.  When $\omega t = 3\pi/2,$ we have $I_1 = -I_{c1}$ and $I_2= -(I-I_{c1})$. 

During the next half cycle, for times obeying $3\pi/2 < \omega t < 5\pi/2$, or since the behavior is periodic with period $T = 2\pi/\omega$, for times obeying $-\pi/2 < \omega t < \pi/2$, we have the following behavior.  At time $\omega t = -\pi/2$ we have $I_1 = -I_{c1}$ and $I_2= -(I-I_{c1})$.  As $I_t(t)$ increases, we again have $E_1 = E_2 = 0$, such that equation (\ref{I1dot})  yields ${\dot I}_1=r_{12} {\dot I}_2$, whose solution is now
\begin{eqnarray}
I_{1}(t)&=&-I_{c1}+\frac{r_{12}I(1+\sin \omega t)}{1+r_{12}},\\
I_{2}(t)&=&-(I-I_{c1})+\frac{I(1+\sin \omega t)}{1+r_{12}}. 
\end{eqnarray}
These equations hold  for times obeying $-\pi/2 \le \omega t \le \omega t_{max}$, where $t_{max}$ is the time that $I_1$ plateaus at its maximum value,  
\begin{equation}
I_{1}(t_{max})=-I_{c1}+\frac{r_{12}I(1+\sin \omega t_{max})}{1+r_{12}}=I_{c1},
\end{equation}
such that 
\begin{equation}
\sin\omega t_{max} = 2F_{ft}/F-1,
\label{omegatmax2nd}
\end{equation}
where $F_{ft}< F < 1$.
For times such that $\omega t_{max} \le \omega t \le \pi/2$, we have $I_1(t) = I_{c1}$ and $I_2(t) = I_t(t) -I_{c1}$.  

Plots of $I_1(t),$ $I_2(t)$, and $I_t(t) = I_1(t)+I_2(t) = I \sin \omega t$ calculated from equations (\ref{Fft2nd})-(\ref{omegatmax2nd}) would look very similar to the plots shown in Fig. \ref{4ab}(a), except that the roles of $I_1$ and $I_2$ would be interchanged.  Similarly, plots of $dI_2/dt$ would look very similar to the plots of $dI_1/dt$ in Fig. \ref{4ab}(b).

The ac flux-transfer loss per cycle per unit length of cable is given by 
\begin{equation}
Q_{ft}' = \int_0^T V_2'(t)I_t(t) dt, 
\end{equation}
where $T = 2\pi/\omega$ is the period of one cycle.  Since $E_2 = 0$ at all times and $I_1 = I_t - I_2$, we have
\begin{equation}
V_2 ' = (L_{22}'-L_{21}')dI_2/dt + L_{21}'dI_t/dt.
\end{equation}
However, since the second term on the right-hand side is a purely inductive contribution, it does not contribute to the ac losses.  Thus $Q_{ft}'$ is proportional to the time integral over one period of the product of $dI_2/dt$ and $I_t(t) = I \sin \omega t.$  The result for $r_{12} > r_c$ can be expressed as 
\begin{equation}
Q_{ft}' = 4(L_{22}'-L_{21}')I_{c} I_{c1} (F-F_{ft}),
\label{Qftr12large}
\end{equation}
where $F_{ft}$ is given by equation (\ref{Fft2nd}).
Equation (\ref{Qftr12large}) also can be obtained by noting that energy is dissipated only over those portions of the cycle for which $E_1 \ne 0$, $|I_1| \approx I_{c1}$, and $E_1 = (L_{22}'-L_{12}'){\dot I}_2$ [see equations (\ref{V1'p}) and (\ref{V2'p})].  The coefficient of $I_{c1}$ on the right-hand side of equation (\ref{Qftr12large}) is twice the magnitude of the magnetic flux per unit length that passes through the superconducting layer 1 during the resistive portion of a half cycle.

\section{Estimates and Conclusions}

Let us now estimate the size of the flux-transfer loss from equation (\ref{Qftr12small}).  Let us first consider a two-layer counter-wound cable with $t_1=-t_2$, $R_1 \approx R_2$, $I_{c2} = I_c/2$, in the limit of large $\chi$.
As already pointed out in our earlier publication \cite{Clem10}, with an {\it in-out} configuration, both $I_1/I_2$ (equation (\ref{I10byI20p})) and $F_{ft}$ (equation (\ref{Fft2nd})) are readily shown to be close to 1, which means that there is no flux-transfer loss over nearly the full range of applied current.  However, for an {\it out-in} configuration (note that we by mistake called this {\it in-out} just before equation (15) of our earlier article \cite{Clem10}) and the same parameters, $I_1/I_2 = t_{22}$; so, for example, at $\theta = 20^\circ$, $F_{ft}$ = 0.566 ($I_{rms}/I_c$ = 0.40).  So for $F>F_{ft}$, the flux-transfer loss from equation (\ref{Qftr12small}) is 
\begin{equation}
Q_{ft} = (\mu_0/\pi R_2) \chi d_2 I_c^2 (F-F_{ft}).
\label{QftE&C}
\end{equation}
This loss, induced by the magnetic layers between the superconductor layers, diverges with $\chi$.  For a frequency $f$ = 60 Hz, $I_c$ = 5000 A, $d_2$ = 150 $\mu$m (assuming an {\it out-in} configuration with two 75 $\mu$m thick substrates), and even a modest $\chi$ = 10, one finds a very substantial loss of $Q_{ft}f = 60 (F - F_{ft})$ W/m.

Consider even a relatively small excess in current above the flux-transfer threshold: $F - F_{ft}$ = 0.1. Equation $(\ref{QftE&C})$ then gives 6 W/m; so even over only a kilometer, one phase would generate a 77 K loss of 6 kW and three phases 18 kW.  Additional superconductor losses would also arise in the screening layer in a typical design \cite{Kalsi11}, and to these losses must be added the thermal losses arising from heat leaking in through the cryostat from the environment, dielectric losses and eddy-current losses in other metallic components of the cable.  Given that these low temperature (77 K) losses must be ejected from the system by refrigeration with a typical coefficient of performance (COP) of no better than 0.1, the actual power load on the system is $>$180 kW.  Thus the flux-transfer losses in this case would constitute a net loss of 
$>$180/48000 or $>$0.4\% in a three-phase cable carrying 2000 A$_{rms}$ at, say, 13.8 kV$_{rms}$ (power = $\sqrt{3}$I$_{rms}$V$_{rms}$ = 48 MVA).  For a single short link in the grid, such a loss is completely unacceptable, being higher than a typical conventional copper cable, even without accounting for all the other losses in the system.  And the cost of installing such a large refrigeration system to handle so much cooling for such a short link is also prohibitive.  Thus it is evident that the large flux-transfer losses must be avoided at all costs.  These considerations highlight the importance of using the {\it in-out} configuration to avoid the large imbalance between the two winding layers and the consequent large flux-transfer losses at intermediate current levels.

Of course, in the absence of magnetic substrates ($\chi$ = 0 in equations (\ref{L11p})-(\ref{L22p})), and with two counter-wound layers ($t_1 = -t_2$) with $R_1 \approx R_2$, both $I_1/I_2$ (equation (\ref{I10byI20p})) and $F_{ft}$ (equation (\ref{Fft2nd}))  are close to 1; so, as in the magnetic {\it in-out} case, the two winding layers are inductively balanced, and there are no flux-transfer losses until the current total amplitude nears the total critical current of the cable.  

The simplicity of the two-layer counter-wound configuration and its minimized losses when magnetic substrates are avoided has created a significant incentive for wire manufacturers to develop wires without magnetic substrates and with sufficiently high critical currents to meet required cable ratings with only two layers.  In fact, recent progress is rapidly attaining these objectives.  Many manufacturers already use a process based on non-magnetic Hastelloy substrates \cite{Selvamanickam11}. A non-magnetic Ni-9W substrate has also been developed to replace the magnetic Ni-5W \cite{Rupich12}.  And critical currents have been significantly enhanced by optimization of pinning and increased film thickness in 2G HTS wires \cite{Rupich12,Selvamanickam11}.   Putting together all the progress in wire capability and enhanced understanding of the origin of AC losses, a single phase cable with a two-layer winding has recently been demonstrated with superconductor AC loss as low as 0.12 W/m at 3000 A$_{rms}$ \cite{Yagi12}! 

\ack{	JRC's work on this paper was supported primarily by AMSC but also in part at Ames Laboratory with support by the Department of Energy - Basic Energy Sciences under Contract No. DE-AC027CH11358  and by the Center for Emergent Superconductivity, an Energy Frontier Research Center funded by the U.S. Department of Energy, Office of Science, Office of Basic Energy Sciences under Award Number DE-AC0298CH1088; his work in establishing the framework for this theory was done as part of a research project funded by Pirelli Cable Corporation (now Prysmian Power Cables and Systems USA LLC).  The authors thank G. Snitchler, S. Fleshler and W. Carter for helpful conversations.
}


\begin{thebibliography}{99}
\bibitem{foot1} References \cite{Garber76}  and \cite{Daeumling99}   presented equations consistent with those in equations (\ref{Vn'1})-(\ref{Lnm'}).  
However, the expressions for the self and mutual inductances in  \cite{Daeumling99} contained a logarithmic dependence upon $D$, the distance from the center of the cable to a cylindrical return lead.  This apparent dependence upon $D$ cancels when differences between inductances are calculated.  The equations in   \cite{Daeumling99} correspond to evaluating the voltage with a measuring circuit in which the voltmeter leads run along the inside of the return lead, rather than on the outside of the cable of radius $R_N$, as assumed here.  
However, as noted in  \cite{Garber76},  the distribution of currents within the cable cannot depend upon the azimuthal field outside $R_N$, which depends only upon the total current  $I_1 + I_2$, and because of equation (\ref{VnEquality}) the dependence upon $D$ cancels out.  The inductances in  \cite{Daeumling99} reduce to those in equation (\ref{Lnm'}) if $D$ is set equal to $R_N$.
\end{thebibliography}
\end{document}